\pdfoutput=1

\documentclass[11pt,twoside,a4paper,cmspaper,final,collab]{cms-tdr}
\def\svnVersion{493393P}\def\cmsCernNoTag{CERN-EP-2018-225}\def\cmsCernDate{\today}\def\cmsMessage{Published in the European Physical Journal C as \href{http://dx.doi.org/10.1140/epjc/s10052-019-6774-8}{\doi{10.1140/epjc/s10052-019-6774-8.}}}

\begin{document}\cmsNoteHeader{FSQ-13-009}

\hyphenation{had-ron-i-za-tion}
\hyphenation{cal-or-i-me-ter}
\hyphenation{de-vices}

\newlength\cmsTabSkip\setlength{\cmsTabSkip}{1ex}
\ifthenelse{\boolean{cms@external}}{\providecommand{\cmsTable}[1]{#1}}{\providecommand{\cmsTable}[1]{\resizebox{\textwidth}{!}{#1}}}
\newcommand{\upsn}{\ensuremath{\PgU\mathrm{(nS)}}\xspace}
\newcommand{\upsum}{\ensuremath{\PgU\mathrm{(sum)}}\xspace}
\newcommand{\gammap}{\ensuremath{\PGg\Pp}\xspace}
\newcommand{\gammapb}{\ensuremath{\PGg\mathrm{Pb}}\xspace}
\newcommand{\wgammap}{\ensuremath{W_{\PGg\Pp}}\xspace}
\providecommand{\STARLIGHT}{\textsc{starlight}\xspace}
\providecommand{\NA}{\ensuremath{\text{---}}}
\cmsNoteHeader{FSQ-13-009}

\title{Measurement of exclusive $\PgU$ photoproduction from protons in $\Pp$Pb collisions at
$\sqrtsNN = 5.02\TeV$}
\author{The CMS Collaboration}
\date{\today}
\abstract{
The exclusive photoproduction of $\upsn$ meson states from protons,
$\gammap \to \upsn\,\Pp$ (with $\mathrm{n}=1,2,3$), is studied
in ultraperipheral $\Pp$Pb collisions at a centre-of-mass energy per nucleon pair of $\sqrtsNN = 5.02\TeV$.
The measurement is performed using the $\upsn \to \mu^+\mu^-$ decay mode, with
data collected by the CMS experiment corresponding to an integrated luminosity of 32.6\nbinv.
Differential cross sections as  functions of the $\upsn$
transverse momentum squared $\pt^2$, and rapidity $y$, are presented.
The $\PgUa$ photoproduction cross section is extracted in the
rapidity range $\abs{y}< 2.2$, which corresponds to photon-proton
centre-of-mass energies in the range $91<\wgammap<826\GeV$.
The data are compared to theoretical predictions based on perturbative quantum
chromodynamics and to previous measurements.
}
\hypersetup{
pdfauthor={CMS Collaboration},
pdftitle={Measurement of exclusive Y photoproduction off protons in pPb collisions at 5.02 TeV},
pdfsubject={CMS},
pdfkeywords={CMS, UPC, photoproduction, Y, pPb}
}
\maketitle

\section{Introduction}
\begin{figure*}[htpb!]
\centering
\includegraphics[width=0.32\textwidth]{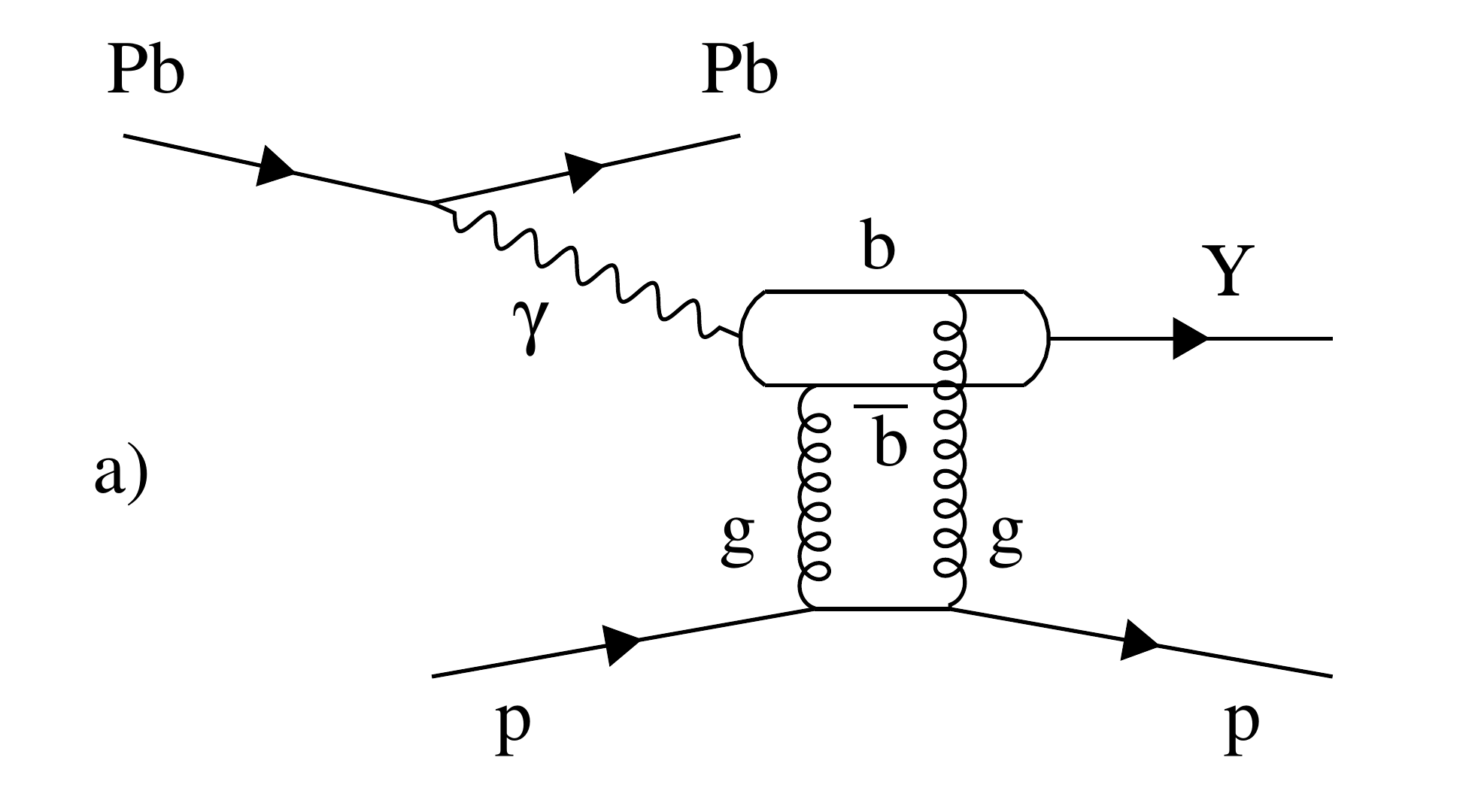}
\includegraphics[width=0.32\textwidth]{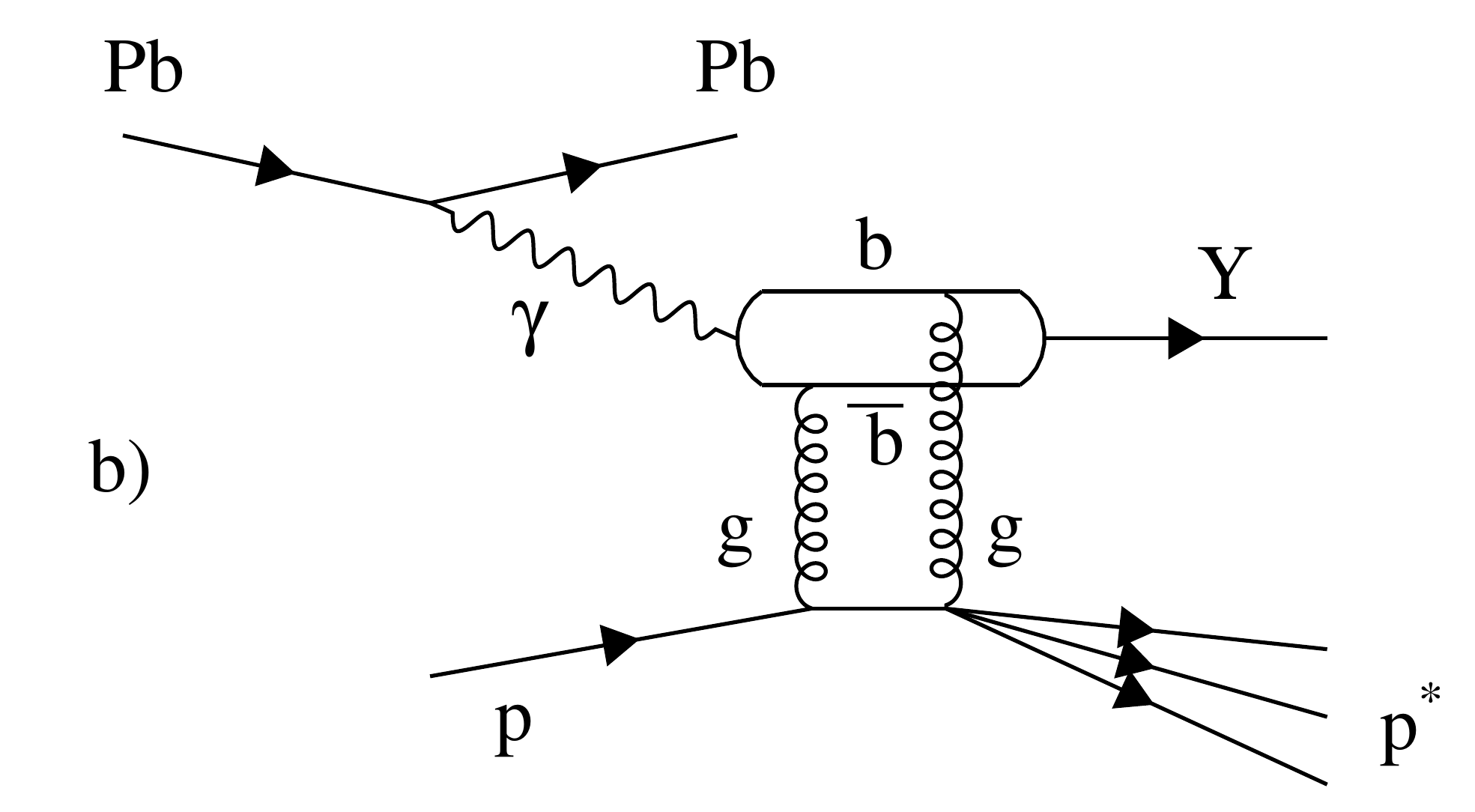}
\includegraphics[width=0.32\textwidth]{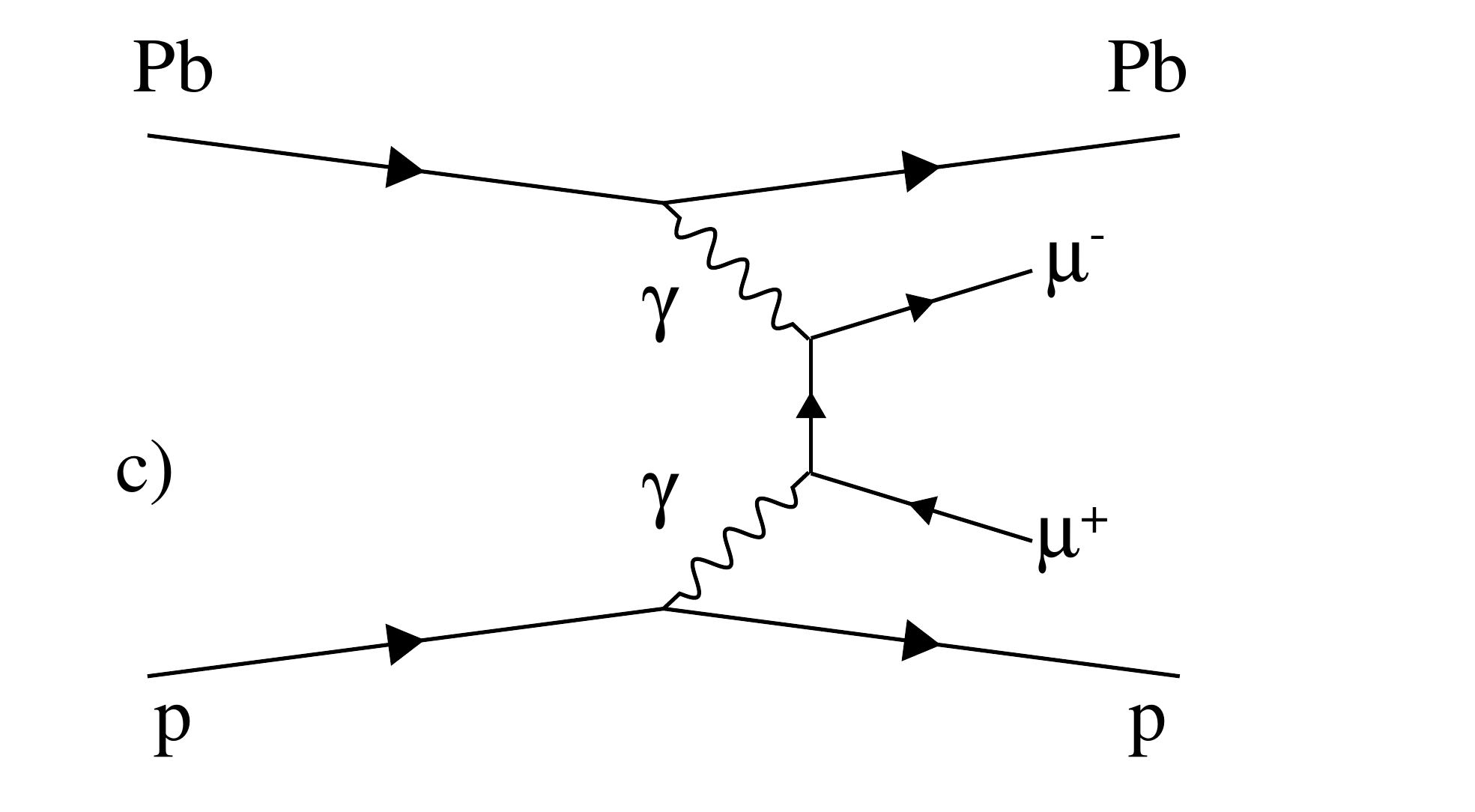}
 \caption{Diagrams representing (a) exclusive $\PgU$ photoproduction, (b) proton dissociative , or ``semiexclusive'',
 $\PgU$ photoproduction, and (c) exclusive dimuon QED continuum production in $\Pp$Pb collisions.}
\label{fig:ex1}
\end{figure*}

This paper reports a first measurement of the exclusive photoproduction of $\PgU$ mesons from protons in $\Pp$Pb collisions at
a nucleon-nucleon centre-of-mass energy of $\sqrtsNN = 5.02\TeV$, performed at the CERN LHC with the CMS detector.
Exclusive photoproduction of vector mesons can be studied at the LHC in ultraperipheral collisions (UPCs) of
protons and/or ions occurring at impact parameters larger than the sum of their radii, thereby
largely suppressing their hadronic interaction~\cite{Baltz:2008bb}. In such UPCs, one of the incoming hadrons
emits a quasi-real photon that converts into a $\qqbar$ (vector meson) bound state following a
colour-singlet gluon exchange with the other ``target'' proton or ion~\cite{Klein:1999qj,d'Enterria:2007pk}.
Since the incoming hadrons remain intact after the interaction and only the vector meson
is produced in the event, the process is called ``exclusive''. Given that the photon flux scales
with the square of the emitting electric charge, the radiation of quasi-real photons from the Pb ion is strongly
enhanced compared to that from the proton. Figure~\ref{fig:ex1}a  shows the dominant diagram for the exclusive
$\PgU$ photoproduction signal in $\Pp$Pb collisions,  $\Pp\mathrm{Pb} \to (\gammap)\mathrm{Pb}\to \Pp\,\PgU\,\mathrm{Pb}$.
If the $\PgU$ photoproduction is followed by the proton breakup, the process is called ``semiexclusive'' (Fig.~\ref{fig:ex1}b).
The exchanged photon can also interact with a photon radiated from the proton~\cite{Baltz:2008bb,Budnev:1975}.
This two-photon collision can produce an exclusive dimuon state, as shown in Fig.~\ref{fig:ex1}c.
Since we are interested in studying exclusive $\PgU$ production via its dimuon decay,
the latter quantum electrodynamics (QED) continuum production constitutes a background
process.

The study of exclusive photoproduction of quarkonia offers a clean probe of
the target hadron structure~\cite{Baltz:2008bb,d'Enterria:2007pk,Lansberg:2008zm}, with the large mass of the
$\cPJgy$ and $\PgU$ mesons providing a hard scale for calculations based on perturbative quantum
chromodynamics (pQCD)~\cite{Brodsky:1994,Frankfurt:1997at,Frankfurt:2001,Ryskin:2008rt}. In the kinematic region studied here,
the photoproduction of $\cPJgy$ and $\PgU$ mesons from protons is sensitive to generalized parton distributions (GPDs),
which can be approximated by the square of the gluon density in the proton~\cite{Brodsky:1994,Frankfurt:1997at,Frankfurt:2001,Ryskin:2008rt,Ryskin:2013jmr,Adeluyi:2013aa,Adeluyi:2013bb,Guzey:2013aa,Guzey:2013bb,Sampaio:2013aa,Machado:2014,Lappi:2010,Lappi:2013,Goncalves:2017}.
Experimentally, exclusive $\cPJgy$ and $\PgU$ photoproduction cross sections have been observed to rise with photon-proton
centre-of-mass energy $\wgammap$, following a power-law dependence $\wgammap^{\delta}$ with
$\delta=0.7$--1.2~\cite{H1:2000,ZEUS:2009}. This reflects the steep rise of the underlying
gluon density in the proton for decreasing values of the momentum fraction $x$ of the proton carried by the struck parton.
The dependence of the exclusive vector meson photoproduction cross section on the squared
four-momentum transfer at the proton vertex $t$,
parameterized at low values of $\abs{t}$ with an exponential function of the form $\exp(-b\abs{t})$~\cite{ZEUS:1998,H1:2000,H1:2013,ZEUS:2012},
has also often been studied; the $b$ slope parameter provides
valuable information on the parton transverse density profile of the proton~\cite{Frankfurt:1997at,Brodsky:1999,Frankfurt:2001}.

{\tolerance=1800 Exclusive $\PgU$ meson photoproduction was first observed in electron-proton collisions
at HERA~\cite{ZEUS:1998,H1:2000,ZEUS:2009,ZEUS:2012} with the quasi-real photon radiated from the electron.
At the CERN LHC, the LHCb~\cite{LHCb:2013aa,LHCb:2014,LHCb:2018}, CMS~\cite{UPC:cmsjpsi},
and ALICE~\cite{ALICE:2013aa,ALICE:2013bb,TheALICE:2014dwa,ALICE:2018} experiments have measured
exclusive photoproduction of $\cPJgy$ mesons in ultraperipheral proton-proton and nuclear collisions.
The LHCb experiment has also reported the measurement of the exclusive $\PgU$ photoproduction cross section
in $\Pp\Pp$ collisions at $\sqrt{s} = 7$ and 8\TeV~\cite{LHCb:2015}. The larger mass of the $\PgU$
meson provides a larger perturbative scale at which the gluon distribution in the proton is sampled,
and thereby reduces theoretical uncertainties in pQCD calculations. This allows the data to
constrain the gluon distributions at low values of
Bjorken $x$ in global PDF fits for the first time~\cite{Jones:2015nna}. The present paper reports the measurement of $\PgU$ photoproduction
in $\Pp$Pb UPCs that probes the gluon density of the proton in the region
$x= m_{\PgU}^2/\wgammap^2= 10^{-4}$--$10^{-2}$~\cite{d'Enterria:2007pk},
where $m_{\PgU}$ is the $\PgU$ meson mass. This CMS  measurement spans a previously
unexplored low-$x$ region between the HERA and LHCb data, and provides additional experimental insights
on the gluon content in the proton. In this low-$x$ regime, nonlinear QCD effects (gluon recombination)
may become important, possibly leading to the saturation of the parton distribution functions
(PDFs)~\cite{Gribov:1983,Mueller:1986,McLerran:1994}.\par}

The measurements presented here are carried out using the $\mu^{+}\mu^{-}$ decays of the $\upsn$ ($\mathrm{n}=1$, 2, 3) bottomonium mesons in the rapidity range $\abs{y}<2.2$ in the laboratory frame. These include differential cross sections as functions of the $\PgU$ rapidity and transverse momentum squared $\pt^2$ (which approximates the absolute value of the four-momentum transfer squared at the proton vertex, $\abs{t}$), as well as the total $\PgUa$ cross section as a function of $\wgammap$. The results are compared to previous measurements and to theoretical predictions based on leading order (LO) and next-to-leading-order (NLO) pQCD calculations~\cite{Ryskin:2013jmr}, as well as on colour dipole~\cite{Sampaio:2013aa,Machado:2014} and gluon saturation~\cite{Lappi:2010,Lappi:2013,Sampaio:2013aa,Machado:2014,Goncalves:2017} approaches.

\section{Experimental setup}

The central feature of the CMS apparatus is a superconducting solenoid
of 6\unit{m} internal diameter, providing a magnetic field of
3.8\unit{T}. Within the solenoid volume are a silicon pixel and strip tracker,
a lead tungstate crystal electromagnetic calorimeter (ECAL), and a brass and scintillator hadron
calorimeter (HCAL), each composed of a barrel and two endcap sections.
The silicon pixel and strip tracker measures charged-particle trajectories within the pseudorapidity
range $\abs{\eta}< 2.5$. It consists of 66 million pixel and 10 million strip sensor elements.
For charged particles with $1 < \pt < 10\GeV$ and $\abs{\eta} < 1.4$, the track resolutions
are typically 1.5\% in \pt~\cite{TRK-11-001}.

Muons are measured in gas-ionisation detectors embedded in the steel flux-return yoke outside
the solenoid over the range $\abs{\eta}< 2.4$, with detection planes based on three technologies:
drift tubes, cathode strip chambers, and resistive-plate chambers.
The reconstruction algorithm considers all tracks in the silicon tracker and
identifies them as muons by looking for compatible signatures in the calorimeters and in the muon system.
Because of the strong magnetic field
and the fine granularity of the tracker, the muon \pt measurement
based on information from the tracker alone has a good resolution~\cite{CMS-PAPER-MUO-10-004}.

Extensive forward calorimetry, based on Cherenkov radiation detectors,
complements the coverage provided by the barrel and endcap calorimeters.
Two hadron forward (HF) calorimeters, consisting of iron absorbers and
embedded radiation-hard quartz fibres, cover $2.9 < \abs{\eta}< 5.2$,
and two zero-degree calorimeters (ZDCs), with alternating layers of tungsten
and quartz fibers, are sensitive to neutrons and photons with $\abs{\eta}> 8.3$~\cite{ZDC:cms}.

The data are collected with a two-level trigger system. The first level of the CMS trigger system, composed of custom hardware processors, uses information from the calorimeters and muon detectors to select the most interesting events~\cite{Khachatryan:2016bia}. The high-level trigger (HLT) processor farm runs a version of the full event reconstruction software optimized for fast processing.
A more detailed description of the CMS detector, together with a definition of the coordinate system used and the relevant kinematic variables, can be found in Ref.~\cite{Chatrchyan:2008zzk}.

\section{Data sample and Monte Carlo simulation}
\label{datasim}

The data set used in this analysis corresponds to 32.6\nbinv of integrated luminosity collected in
$\Pp$Pb collisions by the CMS experiment in 2013, with beam energies of 4\TeV for the protons and 1.58\TeV per nucleon for the lead
nuclei, resulting in a nucleon-nucleon centre-of-mass energy of $\sqrtsNN = 5.02\TeV$.  The data are the sum of
the collected $\Pp$Pb and Pb$\Pp$ collision samples, with the incoming Pb ion going in the $+z$ and $-z$ beam directions,
corresponding to integrated luminosities of 18.8 and 13.8\nbinv, respectively.

{\tolerance=1800 The photon-proton centre-of-mass energy, $\wgammap$, is related to the rapidity $y$ of the
$\PgU$
meson in the laboratory frame by $\wgammap^{2}=2E_\Pp m_{\PgU} \exp(\pm y)$,  where $E_\Pp$
is the proton energy, and the $+(-)$ sign corresponds to the $\Pp$Pb (Pb$\Pp$) beam configuration. This formula, derived neglecting the transverse momenta involved in the interaction, approximates the true value of $\wgammap$ to better than 1 per mille in the $\wgammap$ range of this measurement. The data span the range
$91 < \wgammap < 826\GeV$, with the limits given by the maximum and minimum rapidities, over $\abs{y}<2.2$, of the
$\PgU$ mesons. Because the CMS detector is symmetric along $z$, the $\Pp$Pb and Pb$\Pp$ data samples are merged
in this analysis after changing the sign of $p_{z}$ of the final state particles in the Pb$\Pp$ sample. \par}

The \STARLIGHT (v3.07)~\cite{Starlight:2004,Nystrand:2001jn} Monte Carlo (MC) event generator is used to simulate exclusive $\upsn$
photoproduction events (Fig.~\ref{fig:ex1}a) and the exclusive QED background (Fig.~\ref{fig:ex1}c).
The \STARLIGHT MC assumes that the photon flux from the incoming hadron(s) is described by the Weizs\"a\-cker--Williams
equivalent photon approximation~\cite{Weiz:1934,Will:1934}, and uses an empirical fit of the exclusive
vector meson photoproduction cross sections to the existing HERA $\gammap$ data.
In the $\upsn$ sample, two contributions are simulated, with the photon being emitted either from the
Pb ion or from the proton. The $\gammap$ events where the photon is emitted from the Pb ion constitute the signal,
while the small fraction of $\gamma \mathrm{Pb}$ events with the photon emitted from the proton is treated as a background.
The signal events in the \STARLIGHT MC are
simulated assuming a $\abs{t}$-differential cross section following an $\exp (-b\abs{t})$ dependence,
and a power law dependence of the cross section on the
photon-proton centre-of-mass energy, $\wgammap^{\delta}$, with the exponent $\delta$. In this study,
the $b$ and $\delta$ parameters are tuned to reproduce the data through a reweighting procedure described in Section~\ref{sec:selection}.
The backgrounds from inclusive and semiexclusive $\PgU$ and dimuon production processes are obtained
using templates derived from control samples in the data, as explained in the next section.
All simulated events are passed through the {\GEANTfour}-based~\cite{Geant4:2003ga,Allison:2006ve,Allison:2016lfl} detector simulation
and the event reconstruction chain of CMS.

\section{Event selection and background estimation}
\label{sec:selection}

The $\upsn$ states are studied in their dimuon decay channel. The UPC dimuon events are selected at the trigger
level with a dedicated HLT algorithm, requiring at least one muon and at least one, but not more than six,
tracks in the event. At the offline level, additional selection criteria for muon quality requirements, are applied~\cite{CMS-PAPER-MUO-10-004,CMS_PAS:HIN_13_003}.
In order to minimize the uncertainties related to the low-$\pt$ muon reconstruction inefficiencies, muons with $\pt^{\mu}>3.3\GeV$ are selected in the region $\abs{\eta^{\mu}}<2.2$ in the laboratory frame.
Exclusive events are selected by requiring two opposite-charge muons with a single vertex and no extra charged
particles with $\pt>0.1\GeV$ associated with it.
In addition, no energy deposits in the HF calorimeters are allowed. This is achieved by requiring
that the largest HF tower energy deposit be smaller than 5\GeV.
The HF energy threshold is set to be larger than the detector noise, and is determined from the energy distributions collected in dedicated data taking with no LHC beams.
Furthermore, the rapidity of the muon pair is required to be in the region $\abs{y}<2.2$ in the laboratory frame. Only events with the
$\pt$ of the muon pair between 0.1 and 1\GeV are considered, thereby reducing the contamination from QED pairs at very low $\pt$ and from
$\PgU$ meson production in inclusive and semiexclusive (where the proton dissociates into a low-mass hadronic system,
Fig.~\ref{fig:ex1}b) processes that dominate the region of large dimuon $\pt>1\GeV$.

\begin{figure*}[t!]
\centering
 \includegraphics[width=0.95\textwidth]{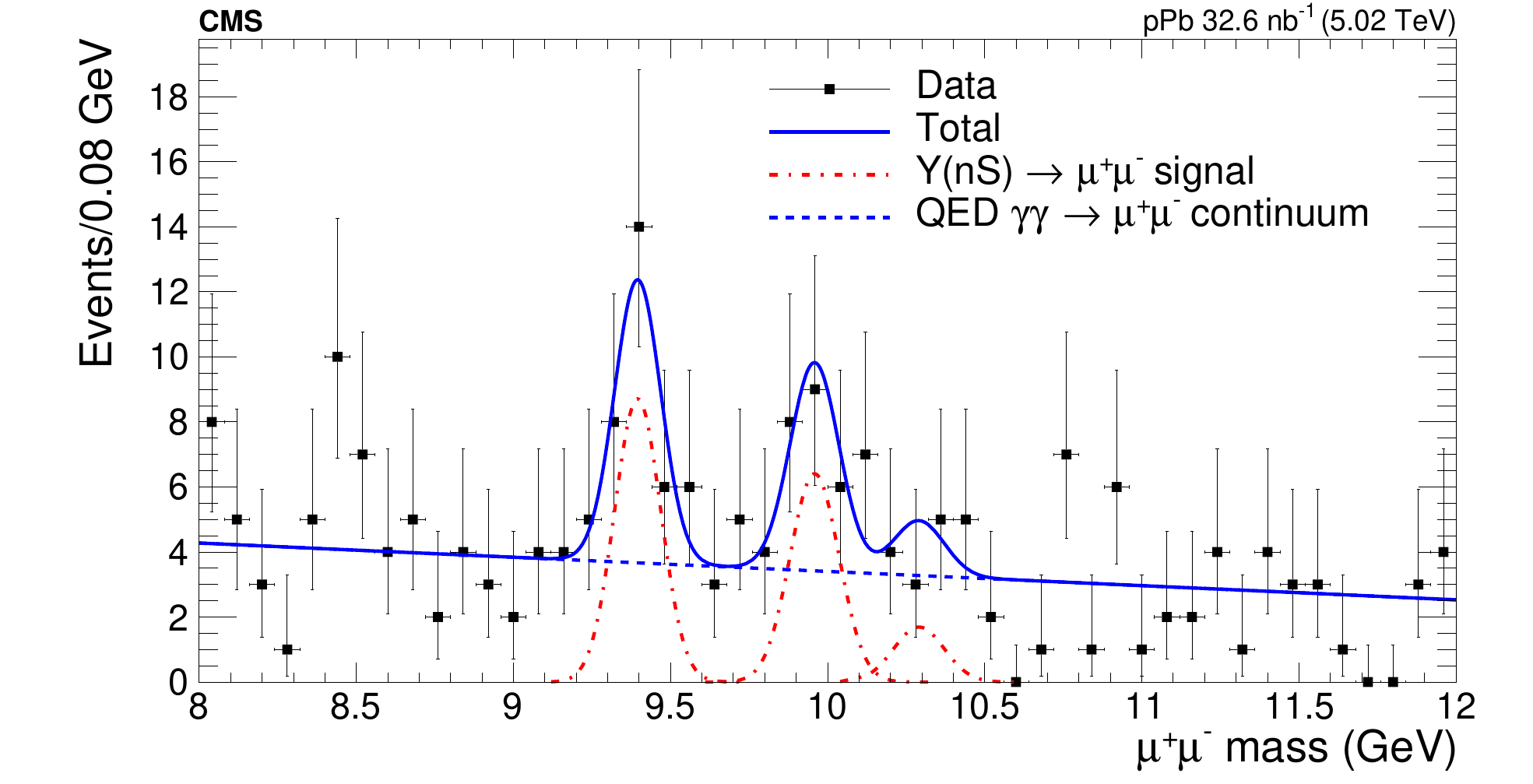}
 \caption{Invariant mass distribution of the exclusive muon pair candidates in the range $8< m_{\mu^{+}\mu^{-}}< 12\GeV$ that pass all the selection criteria, fitted to a linear function for the two-photon QED continuum (blue dashed line) plus three Gaussian distributions corresponding to the $\PgUa$, $\PgUb$, and $\PgUc$ mesons (dashed-dotted-red curves).}
\label{fig:invmass}
\end{figure*}

Figure~\ref{fig:invmass} shows the invariant mass distribution of $\mu^{+}\mu^{-}$ pairs in the range between 8 and 12\GeV
that satisfy the selection
criteria described above. An unbinned likelihood fit to the spectrum is performed using \textsc{RooFit}~\cite{ROOFIT}
with a linear function to describe the QED $\gamma\gamma\to\mu^+\mu^-$ continuum background, where the background slope parameter is fixed to the \STARLIGHT  $\gamma\gamma\to\mu^+\mu^-$ simulation, plus three Gaussian functions for the three $\PgU$ signal peaks, since the natural widths of the $\upsn$ states are much smaller than their
(Gaussian) experimental invariant mass resolutions.
The six free parameters of the fit are the normalizations of the background and the
three signal peaks, as well as the mass and the width of the $\PgUa$ resonance. The $\PgUb-\PgUa$ and
$\PgUc-\PgUa$ mass differences are fixed to their PDG values~\cite{PDG:2016}, while the widths of $\PgUb$ and $\PgUc$
are expressed in terms of the $\PgUa$ width scaled by the ratio of their masses.
The parameters describing the background plus the $\PgUa$
and $\PgUb$ resonances do not change if the $\PgUc$ signal is neglected in the fit.
The statistical significance of the $\PgUa+\PgUb$ peaks over the background is $3.9 \sigma$.
The apparent excess at $8.5\GeV$ has a local significance of $1.6\sigma$, and is consistent with a statistical fluctuation.
Because of the overall small number of events in the data sample,
a determination of the separate $\upsn$ differential cross sections
by fitting the invariant mass spectrum in each $\pt^2$ and $y$ bin leads
to results with large statistical fluctuations.
Instead, the cross sections are extracted by adding up the events, after background subtraction, in the 9.1--10.6\GeV mass region corresponding to the
three $\PgU$ states combined, and the $\PgUa$ yield is derived from the
$\PgUa$/$\upsum$ ratio, where $\upsum=\PgUa+\PgUb+\PgUc$, as described in Section~\ref{sec:xsec}.

\begin{figure}[t!]
\centering
\includegraphics[width=0.49\textwidth]{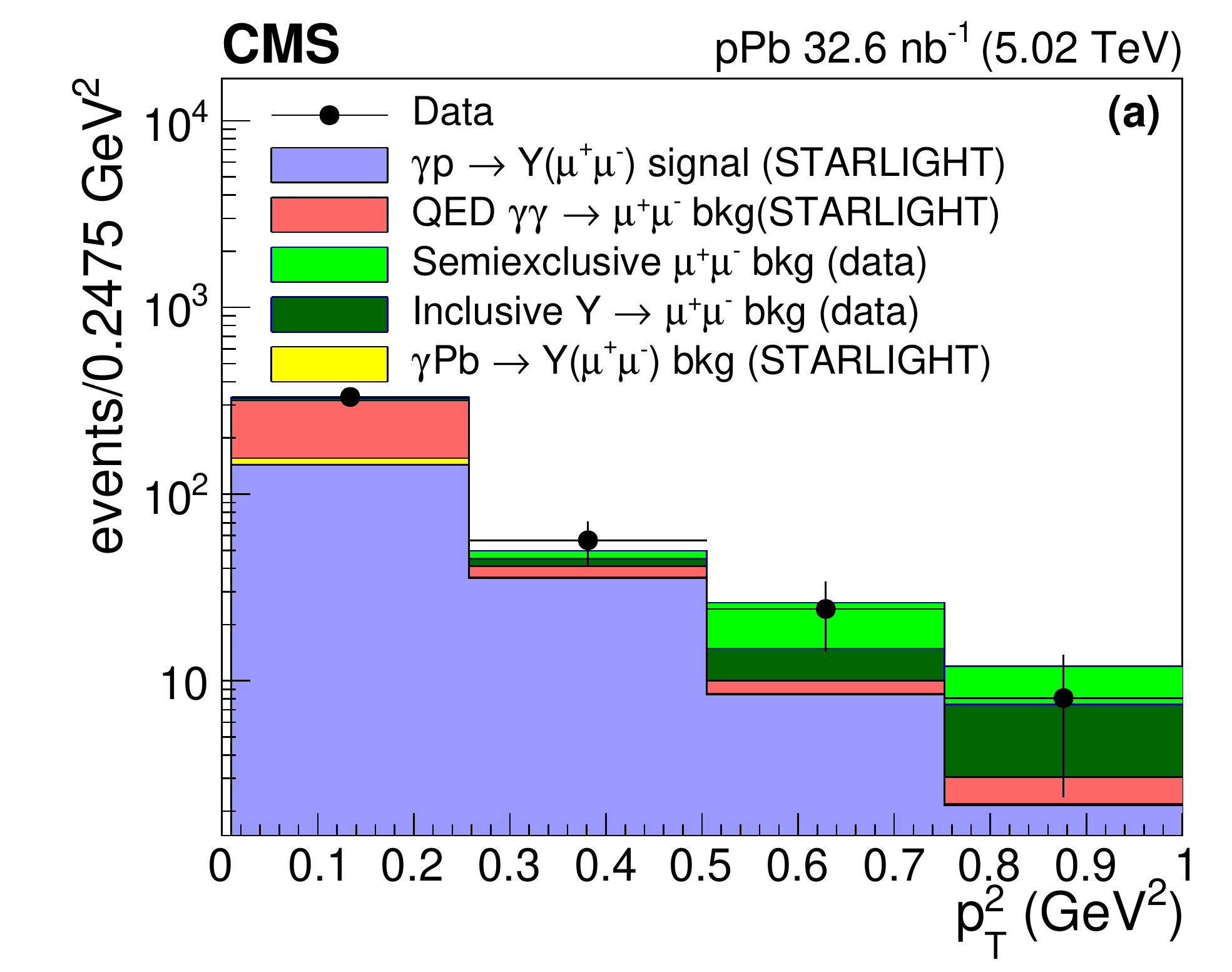}
\includegraphics[width=0.49\textwidth]{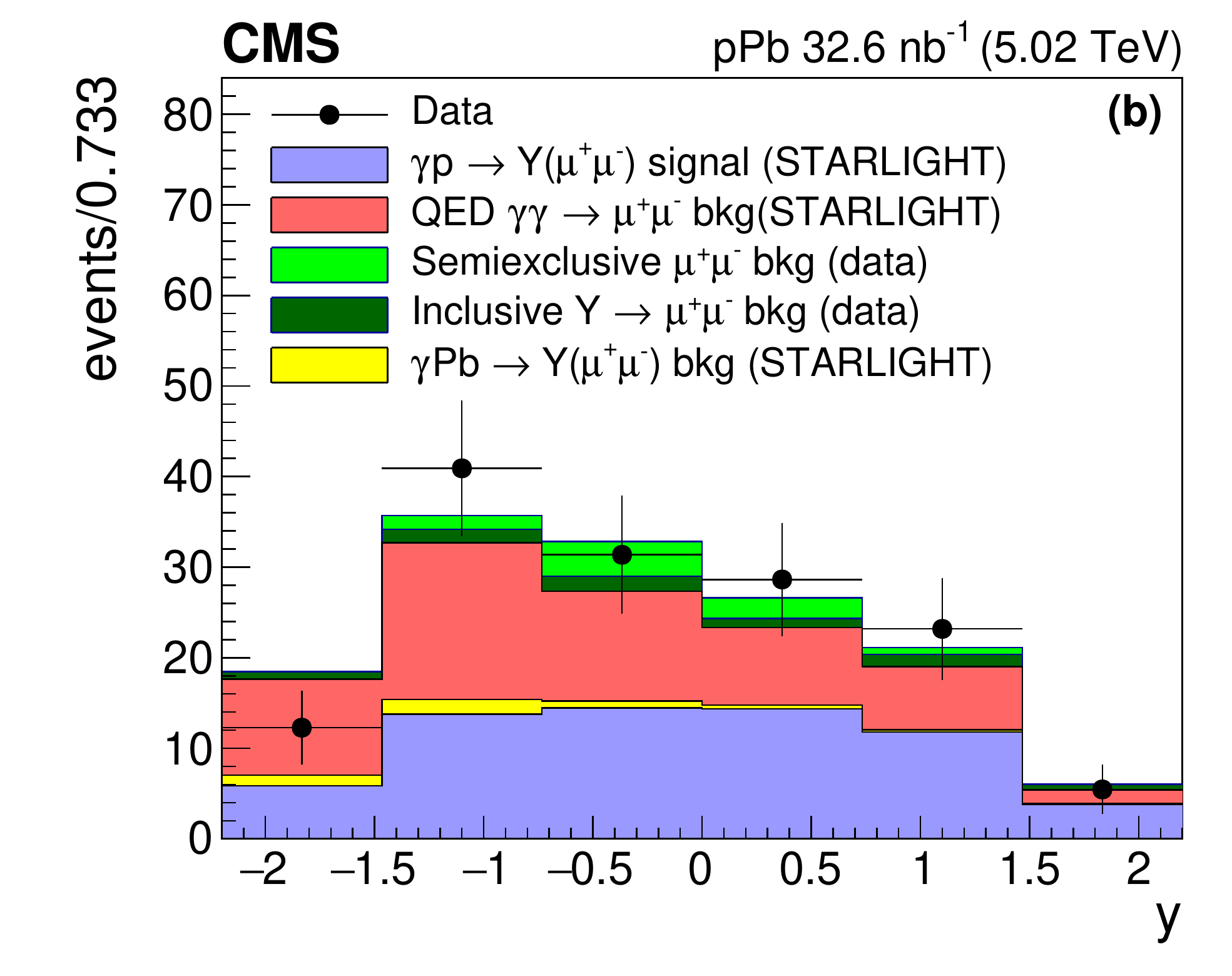}
\caption{Distributions of the (a) transverse momentum squared $\pt^2$,  and (b) rapidity $y$ of exclusive muon pairs with
 invariant mass $9.1<m_{\mu^{+}\mu^{-}}<10.6~\GeV$ after all selection criteria have been applied. Both distributions are compared
to the expectations of signal and background contributions discussed in the text.}
\label{fig:controlplots_pos}
\end{figure}

Figure~\ref{fig:controlplots_pos} shows the dimuon $\pt^2$ and rapidity distributions in
the invariant mass interval $9.1<m_{\mu^{+}\mu^{-}}<10.6\GeV$ for events passing all the selection criteria
for the combined $\Pp$Pb and Pb$\Pp$ samples. The data, uncorrected for detector effects, are compared to the \STARLIGHT
simulation for exclusive $\upsn$ and QED dimuon production, normalized to the recorded integrated luminosity, together with
the inclusive and semiexclusive backgrounds derived from the data themselves as discussed below.
The simulated $\upsn$ events from \STARLIGHT are shown separately for the $\gammap$ and
$\gammapb$ processes;  the latter (with much smaller cross sections) are considered as a background in this analysis. The
$\upsn$ events generated with \STARLIGHT are reweighted to describe the data, using the parameters $b = 5.8\GeV^{-2}$ for the $\abs{t}$
distribution slope, and $\delta=0.99$ for the cross section energy dependence. These parameters minimize
the $\chi^2$ goodness-of-fit value calculated using the data and MC distributions of Fig.~\ref{fig:controlplots_pos}.
The minimization is performed as a function of the rapidity simultaneously for the $\gammap$ and $\gammapb$ samples,
and as a function of $\pt^2$ for the $\gammap$ events.
For $\gammapb$ events, the default \STARLIGHT $\pt$ spectrum is used.

In order to extract the exclusive $\gammap \to \PgU(\mu^+\mu^-) \Pp$  signal events, the exclusive QED and other nonexclusive background contributions need to be subtracted.
The QED $\gamma \gamma \to\mu^+\mu^-$ continuum under the $\upsn$ peaks is estimated with the \STARLIGHT MC simulation.
The absolute prediction of the cross section from this generator is cross-checked by comparing the data and the simulation
in a control region, corresponding to small values of dimuon $\pt$,  $\pt<0.15\GeV$, and  away from the $\PgU$ resonances,
$8<m_{\mu^{+}\mu^{-}}<9.1\GeV$ and $10.6 <m_{\mu^{+}\mu^{-}}<12\GeV$, where the QED process is dominant.
The ratio of the measured yields in the data to those from the \STARLIGHT MC in the control region
is measured to be $1.03 \pm 0.10$, confirming that this event generator reproduces the QED background well,
as observed previously in $\Pp$Pb and PbPb collisions at the LHC~\cite{UPC:cmsjpsi,TheALICE:2014dwa,ALICE:2013aa,ALICE:2013bb}.
The QED contribution, estimated from the \STARLIGHT MC in the signal region,
amounts to 40\% (64 and 8\% in the lowest and highest dimuon $\pt^2$ bins of the corresponding
differential cross section, respectively).

Backgrounds to the exclusive $\PgU\to\mu^+\mu^-$ signal also originate from semiexclusive and
inclusive $\PgU$ meson and Drell--Yan (DY) continuum production,
where any additional hadronic activity falls outside the detector acceptance
 or below the detection thresholds.
These background contributions are
estimated from the data, by removing selectively the neutral or charged exclusivity requirements.
A template dominated by semiexclusive contributions is constructed using events with only
two muon tracks in the tracker accompanied by at least one HF tower having an energy deposit
larger than the noise threshold of 5\GeV, in the direction of the outgoing proton.
Events with two muons satisfying the selection criteria, but with at least one additional track with $\pt>0.1\GeV$,
are used to build a template dominated by inclusive DY production events.
The normalizations of the two templates are obtained from a fit to the measured $\pt^2$ distribution
extended up to $\pt^2=10\GeV^2$, where proton dissociation
and inclusive events dominate, as seen in the tail of the distribution of Fig.~\ref{fig:controlplots_pos}a.
The combination of the normalized inclusive and semiexclusive templates describes the region of
high dimuon $\pt^2$ well in the data in all four $y$ bins used for the cross section extraction.
The overall fraction of both backgrounds in the signal sample
is estimated to be 11\% (3 and 48\% in the lowest and highest dimuon $\pt^2$ bin, respectively).
As an extra cross check of the nonexclusive background subtraction, the signal extraction is repeated
by requiring in addition no neutron detection in the ZDC calorimeters~\cite{UPC:cmsjpsi}.
The extracted yield of exclusive $\PgU$ candidates at low $\pt$ is found to be consistent with
the nominal results without applying the ZDC veto requirement,
thereby confirming the efficiency of the nonexclusive background rejection.

An additional background in this analysis originates from a small contribution of exclusive $\gammapb \to \PgU \mathrm{Pb}$ events.
It is estimated using the reweighted \STARLIGHT $\PgU$ MC sample, and
amounts to 6\% (16 and 1\% in the lowest and highest dimuon $\pt^2$ bin, respectively) of the $\gammap$ MC signal.
Relative to the data, this contribution amounts to 3\% (5 and 1\% at the lowest and
highest dimuon $\pt^2$ bin, respectively).
These simulation-based fractions are used to subtract the $\gammapb \to \PgU \mathrm{Pb}$ contribution from the data.

\section{Extraction of cross sections}
\label{sec:xsec}

The dimuon events selected as described above are used to determine the differential $\PgU$ photoproduction cross sections
in four bins of $\pt^2$ over $\pt^2=0.01$--$1\GeV{}^2$, and in four bins of $y$ over $\abs{y}<2.2$.
Because of the limited size of the data sample, we first extract the differential cross sections for all $\upsn$ resonances combined.
Then, the total cross section as a function of $\wgammap$ is extracted for the $\PgUa$ state alone,
as described below, and is compared with previous experimental measurements and theoretical predictions.

{\tolerance=1600
The background-subtracted $\pt^2$ and $y$ distributions are first unfolded over the region
$0.01<\pt^2<1\GeV{}^2$, $\abs{y}<2.2$, and muon $\pt^{\mu}>3.3\GeV$, by using the
Bayesian iterative unfolding technique~\cite{D'Agostini:1994zf}, as implemented in the \textsc{RooUnfold}
package~\cite{2011arXiv1105.1160A}, with four iterations.
This procedure corrects for detector effects and data migration
between bins. The response matrix is obtained from the \STARLIGHT $\gammap$ simulation.
The differential cross section $\rd\sigma/\rd\pt^2$ is further extrapolated to the
full range of single-muon $\pt$ by means of an acceptance correction factor
$A^{\text{corr}}=N_{\upsn}(\pt^{\mu}>3.3\GeV$$)/N_{\upsn}(\pt^{\mu}>0)$,
estimated with the \STARLIGHT $\gammap$ simulation.
The measured $\rd\sigma/{\rd}y$ values in each rapidity bin are
also similarly extrapolated down to zero dimuon $\pt$.
The $A^{\text{corr}}\approx 0.6$ factor does not significantly depend on
$\pt^2$ but varies as a function of $y$ as shown later in Table~\ref{tab:crossWgp_3uneqbin}.
The $\pt^2$- and $y$-differential cross sections, multiplied by the dimuon branching fraction, are extracted for the three $\upsn$ states combined as follows,
\begin{equation}
 \begin{aligned}
\sum_\mathrm{n}\mathcal{B}_{\upsn\to\mu^+\mu^-}\,\frac{\rd\sigma_{\upsn}}{\rd\pt^2} & = & \frac{N^{\text{corr}}_{\upsum}}{\mathcal{L} \, \Delta \pt^2},\\
\sum_\mathrm{n}\mathcal{B}_{\upsn\to\mu^+\mu^-}\,\frac{\rd\sigma_{\upsn}}{{\rd}y} & = & \frac{N^{\text{corr}}_{\upsum}}{\mathcal{L} \, \Delta y}.
 \end{aligned}
\end{equation}
Here $N^{\text{corr}}_{\upsum}$ denotes the background-subtracted, unfolded, and acceptance-corrected number
of $\PgUa$, $\PgUb$ and $\PgUc$ signal events in each $\pt^2$ and $y$ bin, $\mathcal{L}$ is the
integrated luminosity, $\Delta \pt^2$ and $\Delta y$
are the widths of the $\pt^2$ and $y$ bins, and $\mathcal{B}_{\upsn\to\mu^+\mu^-}$ is the dimuon branching fraction~\cite{PDG:2016}.
The differential $\PgUa$ photoproduction cross section $\rd\sigma_{\PgU{\mathrm{(1S)}}}/{\rd}y$ is then extracted via
\ifthenelse{\boolean{cms@external}}{
\begin{multline}
\label{eq:cross_dsigmady}
\frac{\rd\sigma_{\PgUa}}{{\rd}y}=\\
\frac{f_{\PgUa}}{\mathcal{B}_{\PgUa\to\mu^+\mu^-}(1+f_{\text{FD}})}\left[\sum_\mathrm{ n}\mathcal{B}_{\upsn\to\mu^+\mu^-}\frac{\rd\sigma_{\upsn}}{{\rd}y}\right],
\end{multline}
}{
\begin{equation}
\label{eq:cross_dsigmady}
\frac{\rd\sigma_{\PgUa}}{{\rd}y}=\frac{f_{\PgUa}}{\mathcal{B}_{\PgUa\to\mu^+\mu^-}(1+f_{\text{FD}})}\left[\sum_\mathrm{ n}\mathcal{B}_{\upsn\to\mu^+\mu^-}\frac{\rd\sigma_{\upsn}}{{\rd}y}\right],
\end{equation}
}
where the factor $f_{\PgUa}$ is the ratio of $\PgUa$ to $\upsum=\PgUa+\PgUb+\PgUc$ events,
$f_{\text{FD}}$ is the feed-down contribution to the $\PgUa$ events originating from the $\PgUb\to \PgUa + X$ decays (where
$X=\pi^{+}\pi^{-}$ or $\pi^{0}\pi^{0}$), and $\mathcal{B}_{\PgUa\to\mu^+\mu^-} = (2.48\pm 0.05)\%$~\cite{PDG:2016} is the branching
fraction for the dimuon $\PgUa$ meson decay channel.
\par}

{\tolerance=1200
The fraction of $\PgUa$ to $\upsum=\PgUa+\PgUb+\PgUc$ yields is first derived from the event yield ratios
$r_{21}=N_{\PgUb}/N_{\PgUa} = 0.78 \pm 0.31$ and
$r_{31}=N_{\PgUc}/N_{\PgUa} = 0.21 \pm 0.22$ extracted from the invariant mass fit shown in Fig.~\ref{fig:invmass},
giving $f_{\PgUa}=(1+r_{21}+r_{31})^{-1} = 0.50 \pm 0.09$, where the correlation between the two fitted parameters was not taken into account.
Since this fraction has a relatively large statistical uncertainty, we use the value derived from the analysis~\cite{CMS_PAS:HIN_13_003}
of inclusive $\upsn$ meson production instead, which is performed at the same
nucleon-nucleon collision centre-of-mass energy and in a similar
$\PgU$ rapidity range as the current $\Pp$Pb measurement, in which
the fraction is
 expressed as a function of the number of additional charged particles
in the event ($N_{\text{ch}}$) and extrapolated to $N_{\text{ch}}=0$. This procedure yields
$f_{\PgUa}=0.68 \pm 0.04$, consistent within statistical uncertainties with the factor obtained from the current data, as well as with the
$f_{\PgUa}=0.71 \pm 0.03$ and $0.73 \pm 0.05$ values obtained in the measurements
based on proton-(anti)proton data by LHCb~\cite{LHCb:2015} and CDF~\cite{Acosta:2001gv},  at very forward and
central $\PgU$ rapidities, respectively.
\par}

The feed-down contribution is estimated using the MC simulation in the following way:
the initial $\PgUb$ $\pt$ and $y$ distributions are taken from the \STARLIGHT
generator, and their $\PgUa + \pi\pi$ decays, followed by $\PgUa\to \mu^{+}\mu^{-}$ are simulated
with \PYTHIA~6.4~\cite{Sjostrand:2006za}. After applying all selections, the fraction of dimuon events from $\PgUb$ feed-down
is found to be 8\%  of the exclusive signal $\PgUa$ events reconstructed using the \STARLIGHT simulation.
The contribution from feed-down of exclusive $\PGcb$ states is neglected because these
mesons can only be produced in double-pomeron exchange processes (or in pairs, via
$\gamma\gamma\to\PGcb\PGcb$, with very small cross sections), which
have comparatively much smaller yields in proton-nucleus collisions~\cite{Reeves:1997,Khoze:2010}.

Finally, the exclusive $\PgUa$ photoproduction cross section as a function of
$\wgammap$, is obtained from the $\rd\sigma_{\PgUa}/{\rd}y$
cross section via the relation
\begin{equation}
  \sigma_{\gamma \Pp \to \PgUa\Pp}(\wgammap^{2}) = \frac{1}{\Phi}\frac{\rd\sigma_{\PgUa}}{{\rd}y},
  \label{eq:photo_cross}
\end{equation}
in four different rapidity bins, with associated $\wgammap$ intervals, given in Table~\ref{tab:crossWgp_3uneqbin}.
The cross sections are given at the value $W_{0}$, which corresponds to the average rapidity
over a bin, $\langle y\rangle$.
The photon flux $\Phi$ in Eq.~(\ref{eq:photo_cross}), evaluated at  $\langle y\rangle$,
is obtained from the \STARLIGHT simulation and calculated in the impact parameter space requiring the
$\Pp$Pb separation to be larger than the sum of their radii.

\section{Systematic uncertainties}

{\tolerance=1800
The following sources of systematic uncertainty are taken into account in the measurements of
all differential and total $\PgU$ meson production cross sections, as well as for
the extraction of the exponential slope $b$ of the $\pt^2$ spectrum:
\begin{itemize}
\item The muon reconstruction and selection efficiency has three components: the efficiency to find a track in
  the inner tracker, the efficiency to pass the track quality requirements, and the probability to pass the HLT
  selection. These efficiencies are estimated following the ''tag-and-probe'' method~\cite{CMS_PAS:HIN_13_003},
using first a sample of inclusive $\PgUa$ events selected with a trigger that requires two muons
 (to determine track and muon-quality efficiencies), and second a $\PgUa$ event sample
similar to the one used in the nominal analysis, but collected with an independent trigger
 (to determine the trigger efficiency).
 The associated systematic uncertainty is evaluated from the difference in efficiencies
 obtained from the data and simulation,
 and it leads to uncertainties of $10.5\%$, $4.1\%$ and $1.7\%$ for track,
 muon-quality and trigger component, respectively. The overall uncertainty is estimated
by adding the three numbers in quadrature, and leads to an $11\%$ uncertainty
in the normalization of the cross sections, but  no effect on the $b$ slope measurement.
\item To estimate the systematic uncertainty due to the model dependence of the acceptance correction, the parameters
$b$ and $\delta$ of the simulated \STARLIGHT spectra are changed by $\pm$30\% (chosen conservatively by the
uncertainties of the corresponding fits to the data), and the resulting MC distributions are used for the
determination of the extrapolation factor $A^\mathrm{corr}$, the unfolding, and the $\gammapb \to \PgU \mathrm{Pb}$
background subtraction, resulting in 2--3\% changes in the measured observables.
\item The uncertainty due to the unfolding procedure is studied by modifying the number of iterations used for
the Bayesian unfolding from the nominal value of $4$ to $3$ and $5$, resulting in an uncertainty of $1\%$ for
the $\pt^2$ spectrum, $0.2\%$ for the $b$ slope, and no change for the much flatter $\rd\sigma/{\rd}y$
distribution, which has negligible net bin-to-bin migrations.
\item The uncertainty associated with the exclusive QED background contribution is estimated by comparing the \STARLIGHT
simulation to the data in sideband regions of the invariant mass distribution, $8.0<m_{\mu^{+}\mu^{-}}<9.1\GeV$ and
$10.6<m_{\mu^{+}\mu^{-}}<12.0\GeV$, for $\pt<0.15\GeV$. The ratio of the simulation to the
data in that region is found to be unity with a statistical uncertainty of 5\%.
To estimate the uncertainty due to the QED background subtraction, the MC normalization is
scaled by $\pm$5\%, resulting in 3--4\% changes in the experimental observables.
\item The uncertainty in the nonexclusive background contributions is estimated by varying the HF energy threshold
 by $\pm$10\%. The resulting uncertainties of the observables vary between 3\% and 6\%.
\item The uncertainty introduced by the $\PgUb\to \PgUa+X$ decays is estimated by modifying the values of the $b$ and $\delta$ parameters of the $\PgUb$ spectra in the \STARLIGHT MC
to those obtained from the reweighting described in Section~\ref{sec:selection}. This resulted in a $\pm$2\% variation of the $\PgUa$ cross sections. The uncertainty in $f_{\PgUa}= \PgUa/\upsum$
 is 7\%, estimated as the quadratic sum of the uncertainty obtained from the extrapolation discussed in Section~\ref{sec:xsec} and
from the difference between this result and that obtained by LHCb in Ref.~\cite{LHCb:2015}. The latter takes into account possible differences between inclusive and exclusive processes in proton-proton and proton-lead collisions. An additional 2\% uncertainty in the $\PgUa\to\mu^+\mu^-$ branching fraction is taken from the PDG world average~\cite{PDG:2016}. All these uncertainties affect only the $\PgUa$ cross sections.
\item The theoretical uncertainty in the photon flux affects only the total cross section
 $\sigma_{\gammap \to \PgUa\Pp}$ and is estimated by changing the Pb radius by $\pm$0.5\unit{fm},
conservatively covering different estimates of the neutron skin thickness~\cite{Loizides:2017ack}. It amounts to
2, 3, 3, and 9\% in the four $y$ bins, respectively.
The photon flux uncertainty (listed in the bottom row of Table~\ref{tab:crossWgp_3uneqbin})
is larger for higher photon energies as they are dominated by smaller impact parameters.
\item A systematic normalization uncertainty of $\pm$4\% associated with the integrated luminosity~\cite{CMS-PAS-LUM-13-002}
 is assigned to the measurement of differential and total cross sections,
 with no effect on the $b$ slope uncertainty.
\end{itemize}
\par}

\begin{table*}[hbpt!]
\centering
\topcaption{Relative systematic uncertainties in percent in the measurements of $\sum\mathcal{B}_{\upsn\to\mu^+\mu^-}\rd\sigma/\rd\pt^2$,
the exponential $b$ slope of the $\pt^2$ spectrum,
$\sum\mathcal{B}_{\upsn\to\mu^+\mu^-}\rd\sigma/{\rd}y$, $\rd\sigma_{\PgUa}/{\rd}y$,
and $\sigma_{\gammap \to \PgUa\Pp}$.
Individual contributions, as well as total systematic uncertainties added in quadrature are presented.
For the $\pt^2$- and $y$-differential cross sections, the values averaged over all bins are shown.}
\label{tab:Sys}
\cmsTable{
\begin{tabular}{l c c c c c}
\hline
Source & $\sum\mathcal{B}_{\upsn\to\mu^+\mu^-}\rd\sigma/\rd\pt^2$ & $b$ & $\sum\mathcal{B}_{\upsn\to\mu^+\mu^-}\rd\sigma/{\rd}y$ & $\rd\sigma_{\PgUa}/{\rd}y$ & $\sigma_{\gammap \to \PgUa\Pp}$\\\hline
Muon efficiency                      & $\pm$11 & \NA           &$\pm$11 &$\pm$11  &$\pm$11  \\
Acceptance                           & $\pm$3 &$\pm$2      &$\pm$2 &$\pm$2  &$\pm$2  \\
Unfolding                            & $\pm$1 &$\pm$0.2    & \NA      & \NA       & \NA  \\
Exclusive QED background             & $\pm$4 &$\pm$3     &$\pm$4 &$\pm$4  &$\pm$4   \\
Nonexclusive background              & $\pm$3 &$\pm$3     &$\pm$6 &$\pm$6  &$\pm$6   \\
Integrated luminosity                & $\pm$4 &  \NA          &$\pm$4 &$\pm$4  &$\pm$4   \\
Feed-down                            & \NA       & \NA           & \NA        &$\pm$2  &$\pm$2  \\
Branching fraction $\mathcal{B}_{\PgUa}\to\mu^+\mu^-$  & \NA      &\NA      & \NA        &$\pm$2  &$\pm$2  \\
$f_{\PgUa}$ fraction  & \NA       &  \NA          & \NA        &$\pm$7  &$\pm$7 \\
Photon flux $\Phi$                   & \NA       &  \NA          & \NA        & \NA       &$\pm$4 \\[\cmsTabSkip]
Total                                & $\pm$13 &$\pm$5     &$\pm$14 &$\pm$16 &$\pm$16  \\\hline
\end{tabular}
}
\end{table*}

{\tolerance=1200
The summary of the systematic uncertainties for all measurements is presented in Table~\ref{tab:Sys}.
The dominant sources are the muon reconstruction efficiency and
the modeling of the nonexclusive backgrounds. The total uncertainty is calculated by adding in quadrature the
individual contributions, and varies between $\pm$5\% for the $b$ slope to $\pm$16\% for $\sigma_{\gammap \to \PgUa\Pp}$.
Given the limited integrated luminosity available, the measurements are dominated by statistical uncertainties.
\par}

\section{Results}

\subsection{Differential cross section as a function of  \texorpdfstring{$\pt^2$ and $y$}{pt2 and y}}

\begin{figure*}[htpb!]
  \centering
    \includegraphics[width=0.48\textwidth]{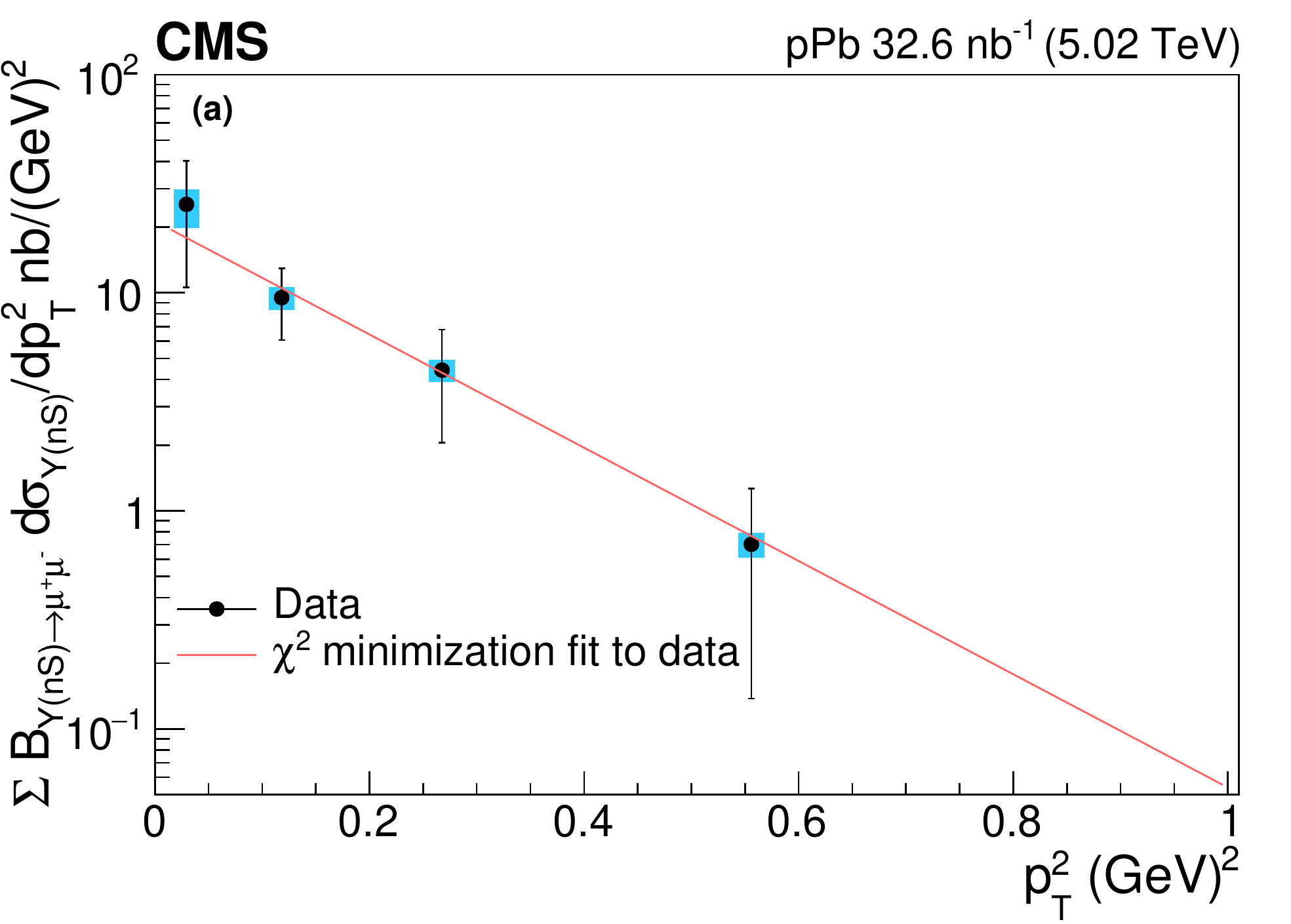}
    \includegraphics[width=0.48\textwidth]{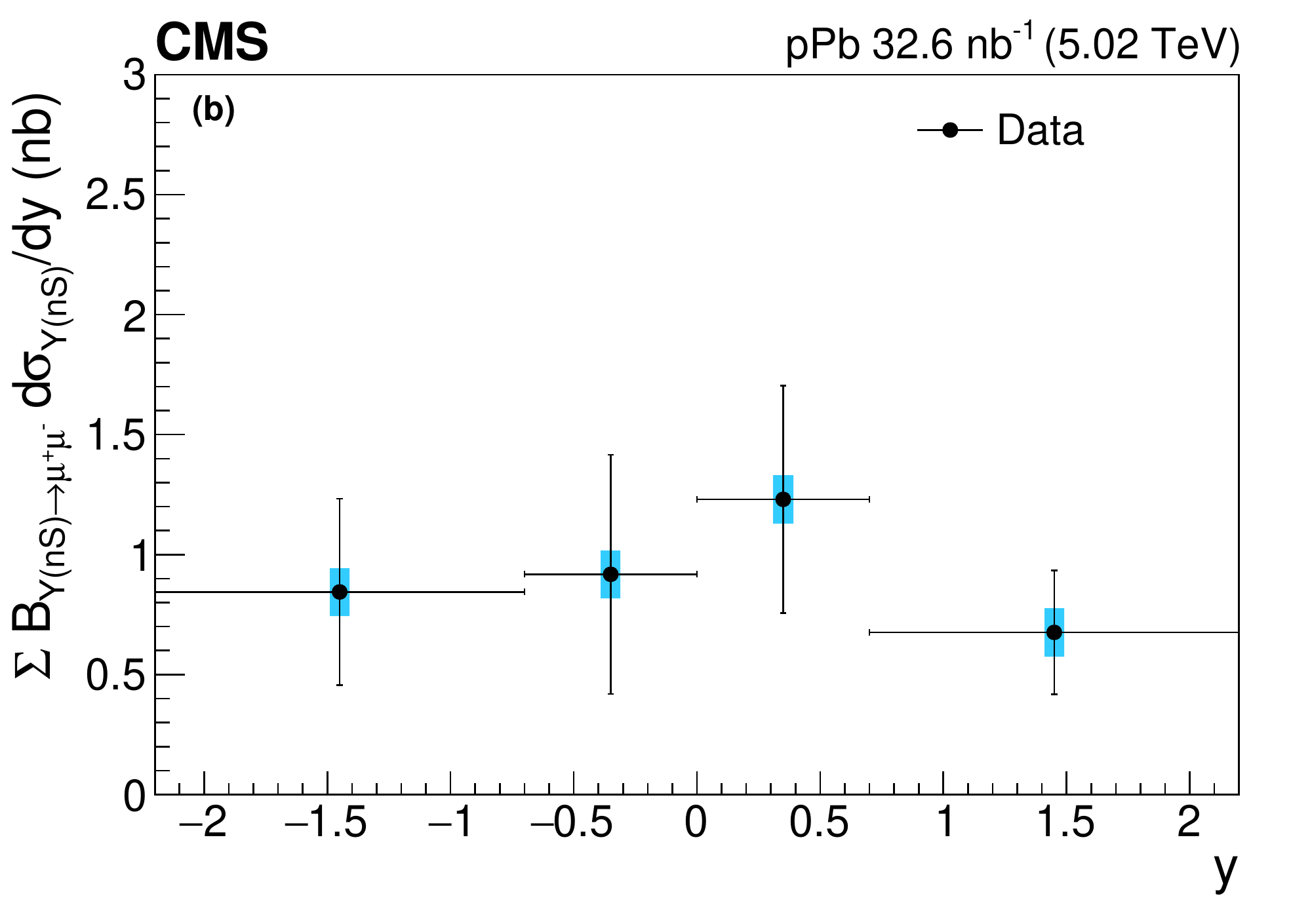}
    \caption{Differential $\upsn\to\mu^+\mu^-$ photoproduction cross section as a function of (a) $\pt^2$
     and (b) rapidity $y$, measured in $\Pp$Pb collisions at $\sqrtsNN = 5.02\TeV$.
     In the left plot, the data points are placed along the abscissa following the prescription of~\cite{LaffertyWyatt},
     and the solid line is an exponential fit of the form $e^{-b\pt^2}$.
     In the right plot, the horizontal bars are shown to indicate the width of each $y$ bin.
     In both plots, the vertical bars represent the statistical uncertainties and the boxes represent the systematic uncertainties.}
    \label{fig:pt2fit}
\end{figure*}

\sloppypar{
The differential cross sections (multiplied by the dimuon branching fractions) for exclusive $\upsn$ photoproduction,
$\sum\mathcal{B}_{\upsn\to\mu^+\mu^-}\rd\sigma_{\upsn}/\rd\pt^2$ and
$\sum\mathcal{B}_{\upsn\to\mu^+\mu^-}\rd\sigma_{\upsn}/{\rd}y$, measured over $\abs{y}<2.2$,
are shown in Fig.~\ref{fig:pt2fit} and tabulated in Table ~\ref{tab:Sys_pt2bin}.
The $\pt^2$-differential cross section is fitted with an exponential function
in the region $0.01<\pt^2<1.0\GeV$$^2$, using a $\chi^2$ goodness-of-fit minimization method.
A slope of $b=6.0 \pm 2.1 \stat \pm 0.3 \syst\GeV^{-2}$ is extracted,
in agreement with the value $b=4.3^{+2.0}_{-1.3} \stat{}^{+0.5}_{-0.6} \syst\GeV^{-2}$
measured by the ZEUS experiment~\cite{ZEUS:2012}
in the photon-proton centre-of-mass energy range $60<\wgammap<220\GeV$, and
with the predictions of pQCD-based models~\cite{Ryskin:2013jmr}.
}

\begin{table*}
\centering
\topcaption{Differential exclusive $\upsn\to\mu^+\mu^-$ photoproduction cross sections in four $\pt^2$ and $y$ bins.
The first and second uncertainties correspond to statistical and systematic components, respectively.}
\label{tab:Sys_pt2bin}
\cmsTable{
\begin{tabular}{c c c c }\hline
$\pt^2$ bin ($\GeVns^2$) & $\sum\mathcal{B}_{\upsn\to\mu^+\mu^-}\,\rd\sigma_{\upsn}/\rd\pt^2$ (nb/$\GeVns^2$) & $y$ bin & $\sum\mathcal{B}_{\upsn\to\mu^+\mu^-}\,\rd\sigma_{\upsn}/{\rd}y$ (nb)
\\ \hline
 (0.01,0.05) &  $25.4 \pm 14.8 \pm 4.9$ & $(-2.2, -0.7)$ & $0.8 \pm 0.4 \pm 0.1$ \\
 (0.05,0.20) &  $9.5 \pm 3.4 \pm 1.1$   & $(-0.7,  0.0)$ & $0.9 \pm 0.5 \pm 0.1$ \\
 (0.20,0.35) &  $4.4 \pm 2.4 \pm 0.5$   & $(0.0,  0.7)$ & $1.2 \pm 0.5 \pm 0.1$ \\
 (0.35,1.00) &  $0.7 \pm 0.6 \pm 0.1$   & $(0.7,  2.2)$ & $0.7 \pm 0.2 \pm 0.1$ \\\hline
\end{tabular}
}
\end{table*}

\begin{figure}[htpb!]
\centering
\includegraphics[width=0.49\textwidth]{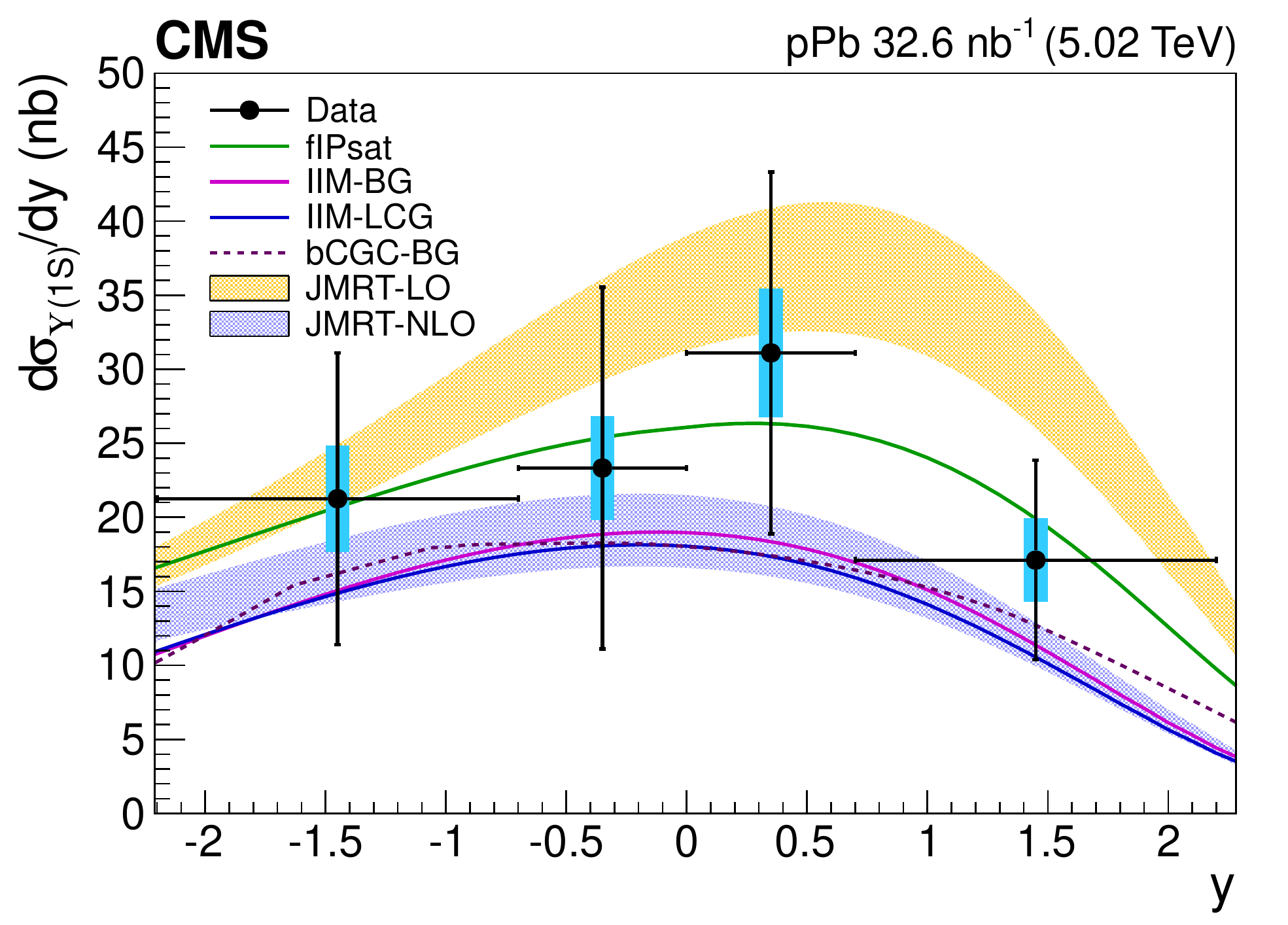}
\caption{Differential $\PgUa$ photoproduction cross section as a function of rapidity measured in $\Pp$Pb
collisions at $\sqrtsNN = 5.02\TeV$ in the dimuon rapidity region $\abs{y}<2.2$,
compared to  various theoretical predictions~\cite{Ryskin:2013jmr,Lappi:2010,Lappi:2013,Sampaio:2013aa,Machado:2014,Goncalves:2017}.
The horizontal bars are plotted to indicate the width of each $y$ bin.
The vertical bars represent the statistical uncertainties and the boxes represent the systematic uncertainties.}
\label{fig:dsigmady_1s}
\end{figure}

Figure~\ref{fig:dsigmady_1s} shows the rapidity distribution of the $\PgUa$ state obtained
according to Eq.~(\ref{eq:cross_dsigmady}). The values of all relevant parameters needed to compute the $\PgUa$ cross sections
in the four rapidity bins under consideration are listed in Table~ \ref{tab:crossWgp_3uneqbin}.
The CMS measurements are compared to the following theoretical predictions:
\begin{itemize}
\item The JMRT model~\cite{Ryskin:2013jmr}, a pQCD approach that uses  standard (collinear) PDFs with a skewness factor to
approximate GPDs, including LO and NLO corrections,
and a gap survival factor to account for the exclusive production;
\item The factorized impact parameter saturation model, fIPsat, with an eikonalized gluon distribution function that uses the colour
  glass condensate (CGC) formalism to incorporate gluon saturation at low $x$~\cite{Lappi:2010,Lappi:2013};
\item the Iancu, Itakura and Munier (IIM) colour dipole formalism~\cite{Iancu:2003ge} with two sets of meson wave functions, boosted Gaussian (BG) and light-cone
  Gaussian (LCG), which also incorporate saturation effects~\cite{Sampaio:2013aa,Machado:2014};
\item the impact parameter CGC model (bCGC), which takes into account the $t$-dependence of the differential
  cross section, using the BG wave function~\cite{Goncalves:2017,Goncalves:2017wgg}.
\end{itemize}
As shown in Fig.~\ref{fig:dsigmady_1s},
most theoretical predictions are consistent with the data, within
the relatively large experimental uncertainties, with the JMRT-LO results being systematically above the data
points as well as all the other calculations.

\begin{table*}[htpb!]
  \centering
    \topcaption{Values of the $\PgUa$ photoproduction cross section in four rapidity $y$ bins,
      corresponding to four photon-proton $\wgammap$ centre-of-mass energy ranges (with central
      $W_{0}$ value obtained following the procedure outlined in Ref.~\cite{LaffertyWyatt}),
      in $\Pp$Pb collisions at $\sqrtsNN = 5.02\TeV$. The symbols $N_{\upsum}^{\text{back-sub}}$, $N_{\upsum}^{\text{unfol}}$, and
      $N_{\upsum}^{\text{corr}} $ represent the numbers of $\upsum=\PgUa+\PgUb+\PgUc$  candidates after background subtraction,
      unfolding, and extrapolation with the $A^{\text{corr}}$ factor, respectively; $N_{\PgUa}$ is the extracted
      number of $\PgUa$ mesons, and $\Phi$ is the theoretical effective photon flux (see text).
      The first (second, if given) uncertainty quoted corresponds to the statistical (systematic) component.}
    \label{tab:crossWgp_3uneqbin}
    \begin{tabular}{c c c c c}
      \hline
      $y$ range & $(-2.2,-0.7)$ & $(-0.7,0.0)$ & $(0.0,0.7)$ & $(0.7,2.2)$ \\
      $\langle y\rangle$ &$-1.45$ & $-0.35$      & $0.35$          & $1.45$  \\[\cmsTabSkip]
      $N_{\upsum}^{\text{back-sub}}$ & $14 \pm 6$  &$ 9 \pm 5  $ &$ 12 \pm 5$  & $12 \pm 5$  \\
      $N_{\upsum}^{\text{unfol}}$    & $19 \pm 9$  &$ 13 \pm 7  $ &$ 17\pm 7$  & $16 \pm 6$ \\
      $A^{\text{corr}}$                      & $0.46 \pm 0.01$  &$ 0.61 \pm 0.01$  & $0.61 \pm 0.01$ & $ 0.50 \pm 0.01$ \\
      $N_{\upsum}^{\text{corr}} $    & $41 \pm 19 \pm 7$  &$ 21 \pm 11 \pm 3$  & $28 \pm 11 \pm 4$ & $33 \pm 13 \pm 5 $ \\
      $N_{\PgUa}=\frac{f_{\PgUa}N_{\upsum}}{(1+f_{\mathrm{FD})}}$ &$26 \pm 12 \pm 4$
  &$13 \pm 7 \pm 2$ &$18 \pm 7 \pm 2$  &$21 \pm 8 \pm 3$ \\
      $\rd\sigma_{\PgUa}/{\rd}y$ (nb) & $21 \pm 10 \pm 4$& $23 \pm 12 \pm 3$ &$31 \pm 12 \pm 4$& $17 \pm 7 \pm 3$\\[\cmsTabSkip]
      $\wgammap$ range (\GeVns) & 91--194 &194--275  &275--390 &390--826 \\
      $W_{0}$ (\GeVns) &133 &231 &328  &568 \\
      Photon flux ($\Phi$) & $102.2 \pm 2.0$ & $68.3 \pm 2.0$ & $46.9 \pm 1.4$  & $17.9 \pm 1.6$ \\
      $\sigma_{\gammap \to \PgUa\Pp}$ (pb) &$208\pm 96 \pm 37$& $343\pm 180 \pm 51$ & $663 \pm 260 \pm 93$ & $956 \pm 376 \pm 162$\\
      \hline
    \end{tabular}
\end{table*}

\subsection{Cross section as a function of \texorpdfstring{$\wgammap$}{W[gamma p]}}

\begin{figure*}[t]
\centering
\includegraphics[width=0.9\textwidth]{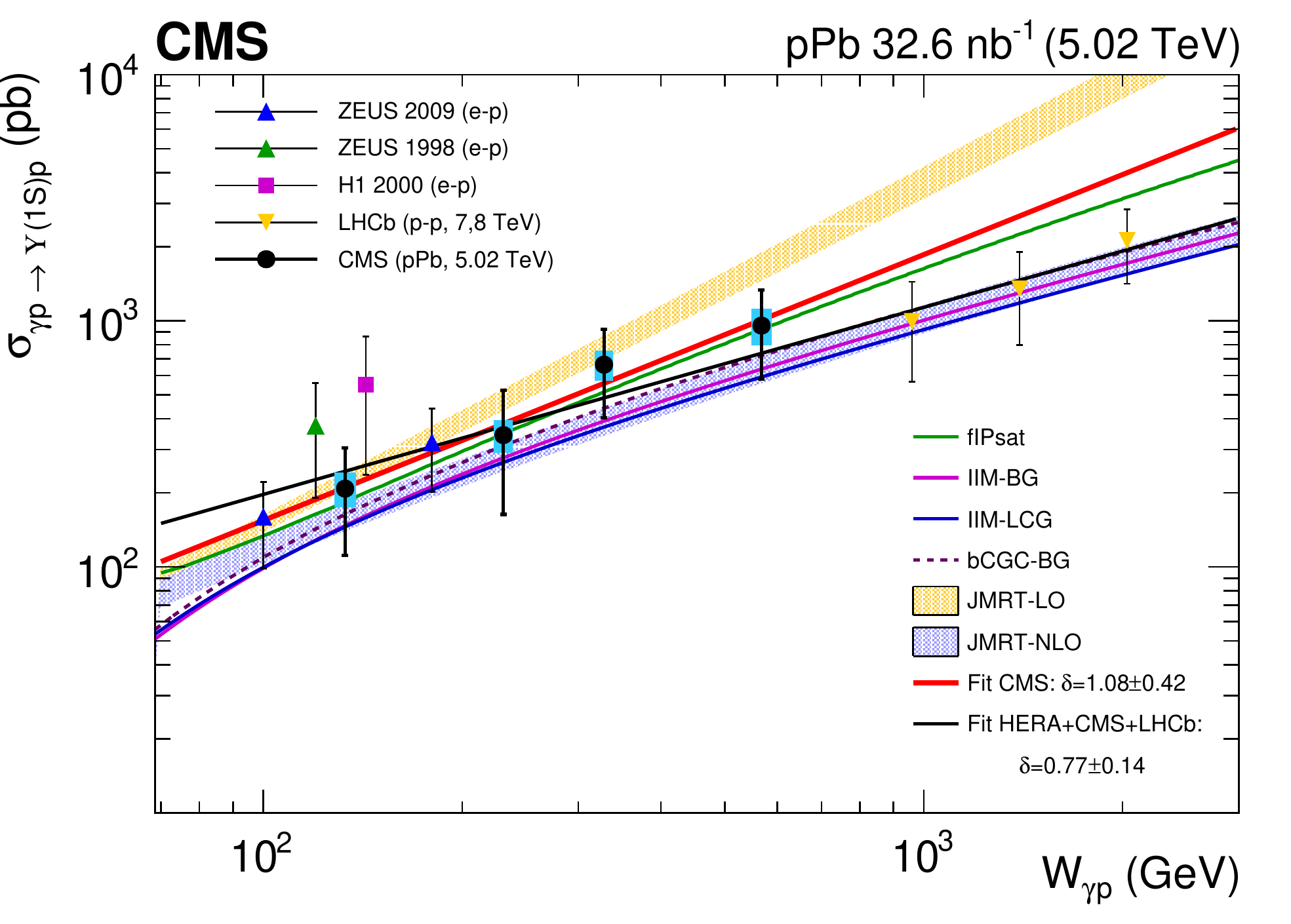}
\caption{Cross section for exclusive $\PgUa$ photoproduction,
$\gammap \to \PgUa \Pp$, as a function of photon-proton centre-of-mass energy, $\wgammap$,
compared to previous HERA~\cite{H1:2000,ZEUS:1998,ZEUS:2009} and LHCb~\cite{LHCb:2015} data
as well as to various theoretical predictions~\cite{Ryskin:2013jmr,Lappi:2010,Lappi:2013,Sampaio:2013aa,Machado:2014,Goncalves:2017}.
 The vertical bars represent the statistical uncertainties and the boxes represent the systematic uncertainties.}
\label{fig:cmsdata_photo_1S_3bin}
\end{figure*}

The values of the $\sigma_{\gammap \to \PgUa\Pp}$ cross section obtained via Eq.~(\ref{eq:photo_cross})
are plotted as a function of $\wgammap$  in Fig.~\ref{fig:cmsdata_photo_1S_3bin},
together with the previous measurements from H1~\cite{H1:2000}, ZEUS~\cite{ZEUS:1998,ZEUS:2009}, and LHCb~\cite{LHCb:2015},
and the five model predictions described in the previous section.
The CMS results (listed in Table~\ref{tab:crossWgp_3uneqbin}) cover the range of energies between the HERA and LHCb data.
As $\sigma(\wgammap)$ is, to first approximation, proportional to the square of the gluon density
of the proton, and since the gluon distribution at low Bjorken $x$ is well described by a power law, the cross section also follows
a power-law energy dependence. A fit  of the extracted CMS $\sigma_{\gammap \to \PgUa\Pp}$ cross section with a
function of the form $A\,(\wgammap[\GeVns]/400)^{\delta}$ (with the constant $A$ corresponding to
the cross section at the middle value, $\wgammap = 400\GeV$, over the range of energies covered)
gives $\delta=1.08\pm 0.42$ and $A=690\pm 183\unit{pb}$ (black solid line in Fig~\ref{fig:cmsdata_photo_1S_3bin}),
consistent with the value $\delta=1.2 \pm 0.8$  obtained by ZEUS~\cite{ZEUS:2009}.
A similar fit to the CMS, H1~\cite{H1:2000}, and ZEUS~\cite{ZEUS:2009} data together gives $\delta=0.99\pm 0.27$, in good
agreement with the results of the fit to the CMS data alone.
The fit over the whole kinematic range, including the higher-$\wgammap$ LHCb data,
yields  an exponent of $\delta=0.77\pm 0.14$, consistent with the collision-energy dependence of the $\cPJgy$ photoproduction
and light vector meson electroproduction cross sections~\cite{Favart:2015umi}.

The data are compared to the predictions of the JMRT model, including LO
and  NLO corrections. A fit with the power-law function in the entire $\wgammap$
range of the data yields $\delta=1.39$ and $\delta=0.84$ for the LO and NLO calculations, respectively. The LO
predictions show a steeper increase of the cross section with energy than seen in the data over the full
kinematic range. The NLO prediction reproduces the measured rise of the cross section with $\wgammap$.
The recent LHCb results at higher $\wgammap$~\cite{LHCb:2015} also disfavour the JMRT LO prediction.
Figure~\ref{fig:cmsdata_photo_1S_3bin} shows theoretical predictions from the fIPsat, IIM, and bCGC models,
which overall bracket the combined HERA and LHC results. The fIPsat calculations are consistent with the CMS measurement,
but predict a somewhat higher cross section than that measured by LHCb. The IIM and bCGC predictions satisfactorily
describe the rise of the cross section with $\gammap$ centre-of-mass energy.
As discussed in Ref.~\cite{Ryskin:2013jmr}, the gluon PDF associated with the JMRT NLO prediction,
which is consistent with the CMS+LHCb data presented here,
has a somewhat different shape at low-$x$ than that predicted by standard pQCD collinear fits used at the LHC such as
CT14~\cite{Dulat:2015mca}, NNPDF3.0~\cite{Ball:2014uwa}, and MMHT~\cite{Harland-Lang:2014zoa}.
However, given the currently large
statistical uncertainty of the results presented here, an improved understanding of the low-$x$ gluon density,
and its evolution with energy scale, will require more precise measurements with larger integrated luminosities and/or at higher centre-of-mass
energies.

\section{Summary}

{\tolerance=2400
The first study of the exclusive photoproduction of $\PgU$(1S,2S,3S) mesons, in the
$\mu^{+}\mu^{-}$ decay mode, from protons in ultraperipheral $\Pp$Pb collisions at $\sqrtsNN = 5.02\TeV$, has been reported
using data collected with the CMS detector corresponding to an integrated luminosity of 32.6\nbinv.
The differential cross section $\rd\sigma/\rd\pt^2$ and associated exponential slope $b$
have been measured in the squared transverse momentum range $\pt^2<1.0\GeV^2$.
The extracted value of $b=6.0 \pm 2.1 \stat \pm 0.3 \syst\GeV^{-2}$ is consistent with the
slope measurement at other centre-of-mass energies.
The exclusive $\PgUa$ photoproduction cross sections, differential in rapidity $y$ and as a function of the
photon-proton centre-of-mass energy $\wgammap$, have been measured in the range $91 < \wgammap< 826\GeV$. Such
measurements probe the region of parton fractional momenta $x\approx 10^{-4}$--$10^{-2}$ in the proton,
bridging a previously unexplored region between the HERA and LHCb measurements.
The dependence of $\sigma_{\gammap\to\PgUa\Pp}$ on $\wgammap$ is well described by a power law with an exponent
smaller than that predicted by leading order perturbative quantum chromodynamics (pQCD) approaches. The exponent is,
however, consistent with that extracted from a fit to the HERA and LHCb data, and with that predicted by
next-to-leading-order pQCD calculations.
The data, within their currently large statistical uncertainties, are consistent with various pQCD
approaches that model the behaviour of the low-$x$ gluon density,
and provide new insights on the gluon distribution in the proton in this poorly explored region.
\par}

\begin{acknowledgments}
We congratulate our colleagues in the CERN accelerator departments for the excellent performance of the LHC and thank the technical and administrative staffs at CERN and at other CMS institutes for their contributions to the success of the CMS effort. In addition, we gratefully acknowledge the computing centres and personnel of the Worldwide LHC Computing Grid for delivering so effectively the computing infrastructure essential to our analyses. Finally, we acknowledge the enduring support for the construction and operation of the LHC and the CMS detector provided by the following funding agencies: BMBWF and FWF (Austria); FNRS and FWO (Belgium); CNPq, CAPES, FAPERJ, FAPERGS, and FAPESP (Brazil); MES (Bulgaria); CERN; CAS, MoST, and NSFC (China); COLCIENCIAS (Colombia); MSES and CSF (Croatia); RPF (Cyprus); SENESCYT (Ecuador); MoER, ERC IUT, and ERDF (Estonia); Academy of Finland, MEC, and HIP (Finland); CEA and CNRS/IN2P3 (France); BMBF, DFG, and HGF (Germany); GSRT (Greece); NKFIA (Hungary); DAE and DST (India); IPM (Iran); SFI (Ireland); INFN (Italy); MSIP and NRF (Republic of Korea); MES (Latvia); LAS (Lithuania); MOE and UM (Malaysia); BUAP, CINVESTAV, CONACYT, LNS, SEP, and UASLP-FAI (Mexico); MOS (Montenegro); MBIE (New Zealand); PAEC (Pakistan); MSHE and NSC (Poland); FCT (Portugal); JINR (Dubna); MON, RosAtom, RAS, RFBR, and NRC KI (Russia); MESTD (Serbia); SEIDI, CPAN, PCTI, and FEDER (Spain); MOSTR (Sri Lanka); Swiss Funding Agencies (Switzerland); MST (Taipei); ThEPCenter, IPST, STAR, and NSTDA (Thailand); TUBITAK and TAEK (Turkey); NASU and SFFR (Ukraine); STFC (United Kingdom); DOE and NSF (USA).

\hyphenation{Rachada-pisek} Individuals have received support from the Marie-Curie programme and the European Research Council and Horizon 2020 Grant, contract No. 675440 (European Union); the Leventis Foundation; the A. P. Sloan Foundation; the Alexander von Humboldt Foundation; the Belgian Federal Science Policy Office; the Fonds pour la Formation \`a la Recherche dans l'Industrie et dans l'Agriculture (FRIA-Belgium); the Agentschap voor Innovatie door Wetenschap en Technologie (IWT-Belgium); the F.R.S.-FNRS and FWO (Belgium) under the ``Excellence of Science - EOS" - be.h project n. 30820817; the Ministry of Education, Youth and Sports (MEYS) of the Czech Republic; the Lend\"ulet (``Momentum") Programme and the J\'anos Bolyai Research Scholarship of the Hungarian Academy of Sciences, the New National Excellence Program \'UNKP, the NKFIA research grants 123842, 123959, 124845, 124850 and 125105 (Hungary); the Council of Science and Industrial Research, India; the HOMING PLUS programme of the Foundation for Polish Science, cofinanced from European Union, Regional Development Fund, the Mobility Plus programme of the Ministry of Science and Higher Education, the National Science Center (Poland), contracts Harmonia 2014/14/M/ST2/00428, Opus 2014/13/B/ST2/02543, 2014/15/B/ST2/03998, and 2015/19/B/ST2/02861, Sonata-bis 2012/07/E/ST2/01406; the National Priorities Research Program by Qatar National Research Fund; the Programa Estatal de Fomento de la Investigaci{\'o}n Cient{\'i}fica y T{\'e}cnica de Excelencia Mar\'{\i}a de Maeztu, grant MDM-2015-0509 and the Programa Severo Ochoa del Principado de Asturias; the Thalis and Aristeia programmes cofinanced by EU-ESF and the Greek NSRF; the Rachadapisek Sompot Fund for Postdoctoral Fellowship, Chulalongkorn University and the Chulalongkorn Academic into Its 2nd Century Project Advancement Project (Thailand); the Welch Foundation, contract C-1845; and the Weston Havens Foundation (USA).
\end{acknowledgments}

\bibliography{auto_generated}

\cleardoublepage \appendix\section{The CMS Collaboration \label{app:collab}}\begin{sloppypar}\hyphenpenalty=5000\widowpenalty=500\clubpenalty=5000\vskip\cmsinstskip
\textbf{Yerevan Physics Institute, Yerevan, Armenia}\\*[0pt]
A.M.~Sirunyan, A.~Tumasyan
\vskip\cmsinstskip
\textbf{Institut f\"{u}r Hochenergiephysik, Wien, Austria}\\*[0pt]
W.~Adam, F.~Ambrogi, E.~Asilar, T.~Bergauer, J.~Brandstetter, E.~Brondolin, M.~Dragicevic, J.~Er\"{o}, A.~Escalante~Del~Valle, M.~Flechl, M.~Friedl, R.~Fr\"{u}hwirth\cmsAuthorMark{1}, V.M.~Ghete, J.~Grossmann, J.~Hrubec, M.~Jeitler\cmsAuthorMark{1}, A.~K\"{o}nig, N.~Krammer, I.~Kr\"{a}tschmer, D.~Liko, T.~Madlener, I.~Mikulec, E.~Pree, N.~Rad, H.~Rohringer, J.~Schieck\cmsAuthorMark{1}, R.~Sch\"{o}fbeck, M.~Spanring, D.~Spitzbart, A.~Taurok, W.~Waltenberger, J.~Wittmann, C.-E.~Wulz\cmsAuthorMark{1}, M.~Zarucki
% ...
\vskip\cmsinstskip
% ...
1:  Also at Vienna University of Technology, Vienna, Austria\\
% ...
\end{sloppypar}
\end{document}